\documentclass[12pt]{article}
\usepackage{graphicx}
\usepackage{latexsym}
\title{QFT as pilot-wave theory of particle creation and destruction}
\author{Hrvoje Nikoli\'c \\
Theoretical Physics Division, Rudjer Bo\v{s}kovi\'{c} Institute, \\
P.O.B. 180, HR-10002 Zagreb, Croatia \\
{\normalsize hrvoje@thphys.irb.hr} \\
\makebox[1in]{} \\
}
\date{\today}
\begin{document}
\maketitle
\begin{abstract}
States in quantum field theory (QFT) are represented by many-particle wave functions,
such that a state describing $n$ particles depends on $n$ spacetime positions.
Since a general state is a superposition of states with different numbers of particles,
the wave function lives in the configuration space identified with a product
of an infinite number of 4-dimensional Minkowski spacetimes. The squared 
absolute value of the wave function is interpreted as the probability density
in the configuration space, from which the standard probabilistic predictions 
of QFT can be recovered. Such a formulation and probabilistic interpretation of QFT
allows to interpret the wave function as a pilot wave that describes deterministic particle trajectories, which automatically includes a deterministic and continuous description
of particle creation and destruction. In particular, when the conditional wave function
associated with a quantum measurement ceases to depend on one of the spacetime
coordinates, then the 4-velocity of the corresponding particle vanishes, describing a 
trajectory that stops at a particular point in spacetime.
In a more general situation a dependence on this spacetime coordinate is negligibly small 
but not strictly zero, in which case the trajectory does not stop but 
the measuring apparatus still behaves as if this particle has been destroyed.
\end{abstract}
\vspace*{0.5cm}

\section{Introduction}

The Bohmian interpretation of nonrelativistic quantum mechanics (QM) 
\cite{bohm1,bohm2,bohmPR1,holbook}
is the best known and most successfull attempt to explain quantum phenomena
in terms of ``hidden variables'', that is, objective properties of the system
that are well defined even in the absence of measurements.
According to this interpretation, particles allways have continuous and deterministic trajectories
in spacetime, while all quantum uncertainties are an artefact of the ignorance of the 
initial particle positions. The wave function plays an auxiliary role, by acting as a 
pilot wave that determines motions of particles for given initial positions.

Yet, in its current form, the Bohmian interpretation
is not without difficulties. An important
nontrivial issue is to make the Bohmian interpretation 
of many-particle systems compatible with special relativity. 
Manifestly relativistic-covariant
Bohmian equations of motion of many-particle systems have been proposed 
in \cite{durr96} and further studied in \cite{nikrel05,nikrel06},
but for a long time it has not been known how to associate probabilistic predictions
with such relativistic-covariant equations of motion.
Recently, a progress has been achieved by realizing that the {\it a priori} probability density
$\rho({\bf x},t) \propto |\psi({\bf x},t)|^2$ of a single particle at the space-position ${\bf x}$ 
at time $t$ should {\em not} be interpreted as a probability density in space satisfying
$\int d^3x \, \rho({\bf x},t)=1$, but as a probability density in {\em spacetime} 
satisfying $\int d^3x \, dt\, \rho({\bf x},t)=1$ \cite{nikprobrel07,nikprobrel09}. 
The usual probabilistic interpretation in space is then recovered as a conditional probability,
corresponding to the case in which the time of detection has been observed.
For an $n$-particle wave function it generalizes to 
$\int d^4x_1 \cdots \int d^4x_n \, \rho(x_1,\cdots , x_n)=1$, where
$\rho(x_1,\cdots , x_n) \propto |\psi(x_1,\cdots , x_n)|^2$ and 
$x_a \equiv ({\bf x}_a, t_a)$. As shown in \cite{nikprobrel07,nikprobrel09},
such a manifestly relativistic-invariant probabilistic interpretation is compatible with
the manifestly relativistic-covariant Bohmian equations of motion.
In this paper we confirm this compatibility, through a more 
careful discussion in Appendix \ref{APPB}.  

The main remaining problem is how to make the Bohmian interpretation
compatible with quantum field theory (QFT), which predicts that particles can 
be created and destructed. How to make a theory that describes deterministic continuous
trajectories compatible with the idea that a trajectory may have a singular point 
at which the trajectory begins or ends? 
One possibility is to explicitly break the rule of continuous deterministic evolution,
by adding an additional equation that specifies stochastic breaking of the trajectories
\cite{durrcr1,durrcr2}. Another possibility is to introduce an additional
continuously and deterministically evolving hidden variable that specifies
effectivity of each particle trajectory \cite{nikcr1,nikcr2}. However,
both possibilities seem rather artificial and contrived. A more elegant possibility
is to replace pointlike particles by extended strings, in which case the Bohmian
equation of motion automatically contains 
a continuous deterministic description of particle creation and destruction 
as string splitting \cite{nikprobrel07}. However, string theory is not yet an 
experimentally confirmed theory, so it would be much more appealing if
Bohmian mechanics could be made consistent without strings.

The purpose of this paper is to generalize the relativistic-covariant Bohmian interpretation
of relativistic QM \cite{nikprobrel09} describing 
a fixed number of particles, to relativistic QFT that describes systems
in which the number of particles may change. It turns out that the
natural Bohmian equations of motion for particles described by QFT automatically
describe their creation and destruction, without need to add any additional structure
to the theory. In particular, the new artificial structures that has been added
in \cite{durrcr1,durrcr2} or \cite{nikcr1,nikcr2} turn out to be completely
unnecessary. 

The main new ideas of this paper are introduced in Sec.~\ref{SEC2} 
in a non-technical and intuitive way. This section serves as a motivation
for styding the technical details developed in the subsequent sections, but
a reader not interested in technical details may be 
satisfied to read this section only. 

Sec.~\ref{SEC3} presents in detail several new
and many not widely known conceptual and technical results
in standard QFT that do not depend on the interpretation of quantum theory.
As such, this section 
may be 
of interest even for readers not interested in interpretations of quantum theory.
The main purpose of this section, however, is to prepare the theoretical framework needed
for the physical interpretation studied in the next section.

Sec.~\ref{SEC4} finally deals with the physical interpretation. 
The general probabilistic interpretation and its relation to the usual probabilistic rules
in practical applications of QFT is discussed first, while 
a detailed discussion of the interpretation in terms
of deterministic particle trajectories (already indicated in Sec.~\ref{SEC2})
is delegated to the final part of this section.

Finally, the conclusions are drawn in Sec.~\ref{SEC5}.

In the paper we use units $\hbar=c=1$ and the metric signature $(+,-,-,-)$.

\section{Main ideas}
\label{SEC2}

In this section we formulate our main ideas in a casual and 
mathematically non-rigorous way, with the intention to develop an
intuitive understanding of our results, and to motivate the formal developments that will
be presented in the subsequent sections.

As a simple example, consider a QFT state of the form 
\begin{equation}\label{e2.1}
|\Psi\rangle = |1\rangle + |2\rangle ,
\end{equation}
which is a superposition of a 1-particle state $|1\rangle$ and a 2-particle state
$|2\rangle$. For example, it may represent an unstable particle for which 
we do not know if it has already decayed into 2 new particles (in which case it is
described by $|2\rangle$) or has not decayed yet (in which case it is
described by $|1\rangle$). However, it is known that
one allways observes either one unstable particle
(the state $|1\rangle$) or two decay products (the state $|2\rangle$). One never 
observes the superposition (\ref{e2.1}). Why?

To answer this question, let us try with a Bohmian approach. 
One can associate a 1-particle wave function
$\Psi_1(x_1)$ with the state $|1\rangle$ and a 2-particle  wave function
$\Psi_2(x_2,x_3)$ with the state $|2\rangle$, where $x_A$ is the spacetime
position $x^{\mu}_A$, $\mu=0,1,2,3$, of the particle labeled by $A=1,2,3$.
Then the state (\ref{e2.1}) is represented by a superposition
\begin{equation}\label{e2.2}
\Psi(x_1,x_2,x_3) = \Psi_1(x_1) + \Psi_2(x_2,x_3) .
\end{equation} 
However, the Bohmian interpretation of such a superposition will describe
{\em three} particle trajectories. On the other hand, we should observe
either one or two particles, not three particles. How to explain that?

To understand it intuitively, we find it instructive to first understand an
analogous but much simpler problem in Bohmian mechanics. Consider
a single non-relativistic particle moving in 3 dimensions. Its wave function 
is $\psi({\bf x})$, where ${\bf x}=\{ x^1,x^2,x^3 \}$ and the
time-dependence is suppressed. Now let us assume that the particle 
can be observed to move either along the $x^1$ direction (in which case it is described
by $\psi_1(x^1)$) or on the $x^2$-$x^3$ plane (in which case it is described
by $\psi_2(x^2,x^3)$). In other words, the particle is observed to move 
either in one dimension or two dimensions, but never in all three dimensions.
But if we do not know which of these two possibilities will be realized, we describe
the system by a superposition
\begin{equation}\label{e2.3}
\psi(x^1,x^2,x^3) = \psi_1(x^1) + \psi_2(x^2,x^3) .
\end{equation}
However, the Bohmian interpretation of the superposition (\ref{e2.3}) 
will lead to a particle that moves in all three dimensions. On the other hand,
we should observe that the particle moves either in one dimension or two dimensions.
The formal analogy with the many-particle problem above is obvious.

Fortunately, it is very well known how to solve this analogous problem involving one
particle that should move either in one or two dimensions. The key is to take into account
the properties of the {\em measuring apparatus}. If it is true that one allways observes
that the particle moves either in one or two dimensions, then the total wave function
describing the entanglement between the measured particle and the measuring apparatus
is not (\ref{e2.3}) but
\begin{equation}\label{e2.4}
\psi({\bf x},y) = \psi_1(x^1)E_1(y) + \psi_2(x^2,x^3)E_2(y) ,
\end{equation}
where $y$ is a position-variable that describes the configuration of the measuring
apparatus. The wave functions $E_1(y)$ and $E_2(y)$ do not overlap.
Hence, if $y$ takes a value $Y$ in the support of $E_2$, then this value is not
in the support of $E_1$, i.e., $E_1(Y)=0$. Consequently, the motion of the
measured particle is described by the conditional wave function \cite{durrequil1}
$\psi_2(x^2,x^3)E_2(Y)$.
The effect is the same as if (\ref{e2.3}) collapsed to $\psi_2(x^2,x^3)$. 

Now essentially the same reasoning can be applied to the superposition (\ref{e2.2}).
If the number of particles is measured, then instead of (\ref{e2.2}) we actually have
a wave function of the form
\begin{equation}\label{e2.5}
\Psi(x_1,x_2,x_3,y) = \Psi_1(x_1)E_1(y) + \Psi_2(x_2,x_3)E_2(y) .
\end{equation}
The detector wave functions $E_1(y)$ and $E_2(y)$ do not overlap. 
Hence, if the particle counter is found in the state $E_2$, then the measured
system originally described by (\ref{e2.2}) is effectively described by $\Psi_2(x_2,x_3)$.

\begin{figure*}[t]
\includegraphics[width=13cm]{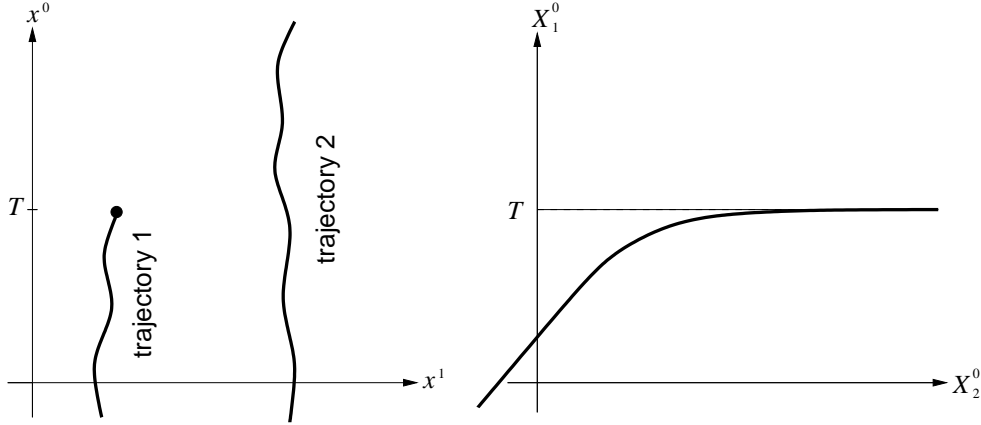}
\caption{\label{fig1}
{\it Left:} The destruction of particle 1 and survival of particle 2, as seen in spacetime.
The dot on trajectory 1 denotes a singular point of destruction at $x^{0}=T$.
{\it Right:} The same process as seen on the $x^{0}_1$-$x^{0}_2$ plane
of the configuration space. Instead of a singular point, we have a continuous 
curve that asymptotically approaches the dashed line $x^{0}_1=T$.}
\end{figure*}

Now, what happens with the particle having the spacetime position $x_1$?
In general, its motion in spacetime may be expected to be
described by the relativistic Bohmian equation
of motion \cite{durr96,nikrel05,nikrel06}
\begin{equation}\label{e2.6}
\frac{dX^{\mu}_1(s)}{ds} = \frac{ 
\frac{i}{2} \Psi^*\!\stackrel{\leftrightarrow\;}{\partial^{\mu}_1}\! \Psi }{\Psi^*\Psi} ,
\end{equation}
where $s$ is an auxiliary scalar parameter along the trajectory. However, in our case
the effective wave function does not depend on $x_1$, i.e., the derivatives in
(\ref{e2.6}) vanish. Consequently, all 4 components of the 4-velocity (\ref{e2.6})
are zero. The particle does not change its spacetime position $X^{\mu}_1$. 
It is an object without an extension not only in space, but also in time. 
It can be thought of as a pointlike particle that exists only at one instant of time
$X^{0}_1$. It lives too short to be detected. Effectively, this particle 
behaves as if it did not exist at all.

Now consider a more realistic variation of the measuring procedure, taking into account
the fact that the measured particles become entangled with the measuring apparatus 
at some finite time $T$.
Before that, the wave function of the measured particles is really well described by 
(\ref{e2.2}). Thus, before the interaction with the measuring apparatus, all 3 
particles described by (\ref{e2.2}) have continuous trajectories in spacetime.
All 3 particles exist. But at time $T$, the total wave function significantly changes.
Either (i) $y$ takes a value from the support of $E_2$ in which case
$dX_1^{\mu}/ds$ becomes zero, or (ii) $y$ takes a value from the support of $E_1$
in which case  $dX_2^{\mu}/ds$ and $dX_3^{\mu}/ds$ become zero.
After time $T$, either the particle 1 does not longer change its spacetime position,
or the particles 2 and 3 do not longer change their spacetime positions. 
The effect is the same as if the particle 1 or the particles 2 and 3 do not exist 
for times $t>T$. In essence, this is how relativistic Bohmian interpretation
describes the particle destruction. In order for this mechanism to work, we see
that it is essential that each particle possesses not only its own space coordinate ${\bf x}_A$,
but also its own time coordinate $x^0_A$.

The corresponding particle trajectories are illustrated by Fig.~\ref{fig1}.
The picture on the left shows the trajectories of particles 1 and 2 in spacetime,
for the case in which the particle 1 is destructed at time $T$. 
The trajectory of the destructed particle looks discontinuous. However, 
the trajectories of all particles are described by continuous functions
$X^{\mu}_A(s)$ with a common parameter $s$,
so the set of all 3 trajectories (because $A=1,2,3$) in the 4-dimensional 
spacetime can be viewed as a single continuous
trajectory in the $3\cdot 4 =12$ dimensional configuration space. 
The picture on the right of Fig.~\ref{fig1}
demonstrates the continuity on the $x^{0}_1$-$x^{0}_2$
plane of the configuration space.

One may object that the mechanism above works only in a very special case
in which the absence of the overlap between $E_1(y)$ and $E_2(y)$ is {\em exact}.
In a more realistic situation this overlap is negligibly small, but not exactly zero.
In such a situation neither of the particles will have exactly zero 4-velocity.
Consequently, neither of the particles will be really destroyed.
Nevertheless, the measuring apparatus will still behave as if some particles 
have been destroyed. For example, if $y$ takes value $Y$ 
for which $E_1(Y)\ll E_2(Y)$, then for all practical purposes the measuring apparatus
behaves as if the wave function collapsed to the second term in (\ref{e2.5}).
The particles with positions $X_2$ and $X_3$ also behave in that way. 
Therefore, even though the particle with the position $X_1$ is not really destroyed,
an effective wave-function collapse still takes place. The influence of the particle
with the position $X_1$ on the measuring apparatus described by $Y$
is negligible, which is effectively the same 
as if this particle has been destroyed.

\begin{figure}[t]
\includegraphics[width=6cm]{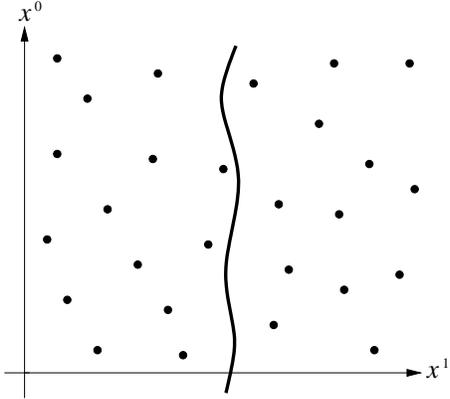}
\caption{\label{fig2}
A ``real'' particle (having non-zero 4-velocity) surrounded by a sea of ``vacuum'' particles
(having zero 4-velocities).}
\end{figure}

Of course, the interaction with the measuring apparatus is not the only mechanism
that may induce destruction of particles. Any interaction with the environment
may do that. (That is why we use the letter $E$ to denote the states of the 
measuring apparatus.) Or more generally, any interactions among particles 
may induce not only particle destruction, but also particle creation. 
Whenever the wave function $\Psi(x_1,x_2,x_3,x_4, \ldots )$ does not really vary
(or when this variation is negligible)
with some of $x_A$ for some range of values of $x_A$,
then at the edge of this range a trajectory of the particle $A$ may exhibit
true (or apparent) creation or destruction. 

In general, a QFT state may be a superposition of $n$-particle states
with $n$ ranging from $0$ to $\infty$. Thus, $\Psi(x_1,x_2,x_3,x_4, \ldots )$
should be viewed as a function that lives in the space of infinitely many coordinates
$x_A$, $A=1,2,3,4,\ldots, \infty$. In particular, the 1-particle wave function
$\Psi_1(x_1)$ should be viewed as a function $\Psi_1(x_1,x_2, \ldots)$
with the property $\partial^{\mu}_A \Psi_1 =0$ for $A=2,3,\ldots, \infty$.
It means that any wave function in QFT describes an infinite number of particles,
even if most of them have zero 4-velocity. As we have already explained, particles
with zero 4-velocity are dots in spacetime. The initial spacetime position of any particle
may take any value, with the probability proportional to
$|\Psi(x_1,x_2, \ldots)|^2$. Thus, the Bohmian particle trajectories associated
with the 1-particle wave function $\Psi_1(x_1,x_2, \ldots)$ take a form as in 
Fig.~\ref{fig2}. In addition to one continuous particle trajectory, there is also an infinite
number of ``vacuum'' particles which live for an infinitesimally short time.

It is intuitively clear that a particle that lives for an infinitesimally short time
is not observable. However, we have an infinite number of such particles, 
so could their overall effect be comparable, or even overwhelming,
with respect to a finite number of ``real'' particles that live for a finite
time? There is a simple intuitive argument that such an effect should
not be expected. The number of ``vacuum'' particles is equal
to the cardinal number of the set on natural numbers, denoted by $\aleph_0$.
This set has a measure zero with respect to a continuous trajectory,
because a continuous trajectory corresponds to a set of real numbers, the cardinal number of
which is $2^{\aleph_0}$ \cite{penrose}. Intuitively, 
the number of points on a single continuous trajectory is infinitely times larger
than the number of points describing the ``vacuum'' particles.
Consequently, the contribution of the ``vacuum'' particles to any 
measurable effect is expected to be negligible.

The purpose of the rest of the paper is to further elaborate
the ideas presented in this section 
and to put them into a more precise framework.

\section{Interpretation-independent aspects of QFT}
\label{SEC3}

\subsection{Measurement in QFT as entanglement with the environment}
\label{SEC3.1}

Let $\{|i\rangle \}$ be some orthonormal basis of 1-particle states.
A general normalized 1-particle state is
\begin{equation}\label{e3.1}
 |\Psi_1\rangle = \sum_i c_i |i\rangle ,
\end{equation}
where the normalization condition implies $ \sum_i |c_i|^2=1$. 
From the basis $\{|i\rangle \}$ one can construct the $n$-particle basis
$\{|i_1,\ldots, i_n\rangle \}$, where
\begin{equation}\label{e3.2}
|i_1,\ldots, i_n\rangle = S_{\{i_1,\ldots, i_n\}}|i_1\rangle \cdots  |i_n\rangle .
\end{equation}
Here $S_{\{i_1,\ldots, i_n\}}$ denotes the symmetrization over all $\{i_1,\ldots, i_n\}$
for bosons, or antisymmetrization for fermions. The most general state in QFT
describing these particles can be written as
\begin{equation}\label{e3.3}
 |\Psi\rangle = c_0 |0\rangle +
\sum_{n=1}^{\infty} \sum_{i_1,\ldots, i_n} c_{n;i_1,\ldots,i_n} 
|i_1,\ldots, i_n\rangle ,
\end{equation}
where the vacuum $|0\rangle$ is also introduced. 
Now the normalization condition implies
$|c_0|^2+\sum_{n=1}^{\infty} \sum_{i_1,\ldots, i_n} |c_{n;i_1,\ldots,i_n}|^2 =1$.

Now let as assume that the number of particles is measured. It implies that the particles
become entangled with the environment, such that the total state describing both 
the measured particles and the environment takes the form
\begin{eqnarray}\label{e3.4}
 |\Psi\rangle_{\rm total} & = & c_0 |0\rangle |E_0\rangle 
\\
& & + \sum_{n=1}^{\infty} \sum_{i_1,\ldots, i_n} c_{n;i_1,\ldots,i_n}  
|i_1,\ldots, i_n\rangle |E_{n;i_1,\ldots,i_n} \rangle .
\nonumber 
\end{eqnarray}
The environment states $|E_0\rangle$, $|E_{n;i_1,\ldots,i_n} \rangle$ are
macroscopically distinct. They describe what the observers really observe.
When an observer observes that the environment is in the state
$|E_0\rangle$ or $|E_{n;i_1,\ldots,i_n} \rangle$, then one says that the 
original measured QFT state is in the state $|0\rangle$ or 
$|i_1,\ldots, i_n\rangle$, respectively. In particular, this is how the number of 
particles is measured in a state (\ref{e3.3}) with an uncertain number of particles.
The probability that the environment will be found in the state
$|E_0\rangle$ or $|E_{n;i_1,\ldots,i_n} \rangle$ is 
equal to $|c_0|^2$ or $|c_{n;i_1,\ldots,i_n}|^2$, respectively.

Of course, (\ref{e3.3}) is not the only way the state $|\Psi\rangle$ can be expanded.
In general, it can be expanded as
\begin{equation}\label{e3.5}
 |\Psi\rangle = \sum_{\xi} c_{\xi} |\xi\rangle ,
\end{equation}
where $|\xi\rangle$ are some normalized (not necessarily orthogonal)
states that do not need to have a definite number of particles.
A particularly important example are coherent states (see, e.g., \cite{bal}),
which minimize the products of uncertainties of fields and their 
canonical momenta. Each coherent state is a superposition
of states with all possible numbers of particles, including zero. 
The coherent states are overcomplete and not orthogonal. Yet, the
expansion (\ref{e3.5}) may be an expansion in terms of coherent states $|\xi\rangle$
as well. 

Furthermore, the entanglement with the environment does not
necessarily need to take the form (\ref{e3.4}). Instead, it may take a
more general form
\begin{equation}\label{e3.6}
 |\Psi\rangle_{\rm total}  = \sum_{\xi} c_{\xi} |\xi\rangle |E_\xi\rangle , 
\end{equation}
where $|E_\xi\rangle$ are macroscopically distinct. 
In principle, the interaction with the environment may create the entanglement
(\ref{e3.6}) with respect to any set of states $\{ |\xi\rangle \}$.
In practice, however, some types of expansions are preferred.
This fact can be explained by the theory of decoherence \cite{schloss}, which 
explains why states of the form of (\ref{e3.6}) are stable only for some particular
sets $\{ |\xi\rangle \}$. In fact, depending on details of the interactions
with the environment, in most real situations the entanglement takes either the form
(\ref{e3.4}) or the form (\ref{e3.6}) with coherent states $|\xi\rangle$.
Since coherent states minimize the uncertainties of fields and their canonical momenta,
they behave very much like classical fields. This explains why experiments in quantum
optics can often be better described in terms of fields rather than particles 
(see, e.g., \cite{bal}). In fact, the theory of decoherence can explain 
under what conditions the coherent-state basis becomes preferred over
basis with definite numbers of particles \cite{zeh,zurek}. 

There is one additional physically interesting class of sets $\{ |\xi\rangle \}$.
They may be eigenstates of the particle number operator defined with respect to
{\em Bogoliubov transformed} (see, e.g., \cite{bd}) creation and destruction operators. 
Thus, even the vacuum may have a nontrivial expansion of the form
\begin{equation}\label{u3.3}
 |0\rangle = c_0 |0'\rangle +
\sum_{n=1}^{\infty} \sum_{i'_1,\ldots, i'_n} c_{n;i'_1,\ldots,i'_n} 
|i'_1,\ldots, i'_n\rangle ,
\end{equation}
where the prime denotes the $n$-particle states with respect to the Bogoliubov 
transformed number operator. In fact, whenever the two definitions of particles
are related by a Bogoliubov transformation, the vacuum for one definition of particles is a 
squeezed state when expressed in terms of particles of the other definition of particles 
\cite{grish}. Thus, if the entanglement with the environment takes the form
\begin{eqnarray}\label{u3.4}
 |\Psi\rangle_{\rm total} & = & c_0 |0'\rangle |E_0\rangle 
\\
& & + \sum_{n=1}^{\infty} \sum_{i'_1,\ldots, i'_n} c_{n;i'_1,\ldots,i'_n}  
|i'_1,\ldots, i'_n\rangle |E_{n;i'_1,\ldots,i'_n} \rangle ,
\nonumber 
\end{eqnarray} 
then the vacuum (\ref{u3.3}) may appear as a state with many particles.
Indeed, this is expected to occur when the particle detector is accelerated 
or when a gravitational field is present \cite{bd}. The theory of decoherence 
can explain why the interaction with the environment leads to an entanglement of the 
form of (\ref{u3.4}) \cite{decohunruh1,decohunruh2,decohunruh3}.

Thus, decoherence induced by interaction with the environment
can explain why do we observe either a definite number
of particles or coherent states that behave very much like classical fields.
However, decoherence alone cannot explain why do we observe 
some particular state of definite number of particles and not some other, 
or why do we observe some particular coherent state and not some other.

\subsection{Free scalar QFT in the particle-position picture}
\label{SEC3.2}

Consider a free scalar hermitian field operator $\hat{\phi}(x)$ satisfying the 
Klein-Gordon equation
\begin{equation}\label{e3.7}
 \partial^{\mu}\partial_{\mu}\hat{\phi}(x)+ m^2\hat{\phi}(x) =0. 
\end{equation}
The field can be decomposed as
\begin{equation}\label{e3.8}
 \hat{\phi}(x)=\hat{\psi}(x)+\hat{\psi}^{\dagger}(x) ,
\end{equation}
where $\hat{\psi}$ and $\hat{\psi}^{\dagger}$ can be expanded as
\begin{eqnarray}\label{e3.9}
& \hat{\psi}(x)=\displaystyle\int d^3k \, f({\bf k}) \, \hat{a}({\bf k})
e^{-i[\omega({\bf k})x^0-{\bf k}{\bf x}]} , &
\nonumber \\
& \hat{\psi}^{\dagger}(x)=\displaystyle\int d^3k \, f({\bf k}) \, \hat{a}^{\dagger}({\bf k})
e^{i[\omega({\bf k})x^0-{\bf k}{\bf x}]} . &
\end{eqnarray}
Here 
\begin{equation}\label{e3.9'}
\omega({\bf k})=\sqrt{{\bf k}^2+m^2}
\end{equation} 
is the $k^0$ component
of the 4-vector $k=\{ k^{\mu} \}$, and
$\hat{a}^{\dagger}({\bf k})$ and $\hat{a}({\bf k})$
are the usual creation and destruction operators, respectively. The function
$f({\bf k})$ is a real positive function which we do not specify explicitly
because several different choices appear in the literature, corresponding to
several different choices of normalization. All subsequent equations will
be written in forms that do not depend on this choice.

We define the operator
\begin{equation}\label{e3.10}
 \hat{\psi}_n(x_{n,1}, \ldots , x_{n,n}) = d_n 
S_{ \{x_{n,1}, \ldots , x_{n,n} \} } 
\hat{\psi}(x_{n,1}) \cdots \hat{\psi}(x_{n,n}) .
\end{equation}
The symbol $ S_{ \{x_{n,1}, \ldots , x_{n,n} \} }$ denotes the symmetrization, 
reminding us that the expression
is symmetric under the exchange of coordinates $\{x_{n,1}, \ldots , x_{n,n} \}$.
(Note, however, that the product of operators on the right hand side of (\ref{e3.10})
is in fact automatically symmetric because 
the operators $\hat{\psi}(x)$ commute, i.e., $[\hat{\psi}(x),\hat{\psi}(x')]=0$.) 
The parameter $d_n$ is a normalization constant determined by the normalization condition
that will be specified below.
The operator (\ref{e3.10}) allows us to define $n$-particle states
in the basis of particle spacetime positions, as
\begin{equation}\label{e3.11}
 |x_{n,1}, \ldots , x_{n,n}\rangle = \hat{\psi}^{\dagger}_n(x_{n,1}, \ldots , x_{n,n})
|0\rangle .
\end{equation}
All states of the form (\ref{e3.11}), together with the vacuum $|0\rangle$, form
a complete and orthogonal basis in the Hilbert space of physical states.

If $|\Psi_n\rangle$ is an arbitrary (but normalized) $n$-particle state,
then this state can be represented by the $n$-particle wave function
\begin{equation}\label{e3.12}
 \psi_n(x_{n,1}, \ldots , x_{n,n}) = \langle x_{n,1}, \ldots , x_{n,n} |\Psi_n\rangle .
\end{equation}
We also have
\begin{equation}\label{e3.12.1}
 \langle x_{n,1}, \ldots , x_{n,n} |\Psi_{n'}\rangle =0 \;\;{\rm for}\;\; n\neq n' .
\end{equation}
We choose the normalization constant $d_n$ in (\ref{e3.10}) such that the following
normalization condition is satisfied
\begin{equation}\label{e3.13}
 \int d^4x_{n,1}\cdots \int d^4x_{n,n} \, | \psi_n(x_{n,1}, \ldots , x_{n,n})|^2 =1 .
\end{equation}
However, this implies that the wave functions 
$\psi_n(x_{n,1}, \ldots , x_{n,n})$ and $\psi_{n'}(x_{n',1}, \ldots , x_{n',n'})$,
with different values of $n$ and $n'$,
are normalized in different spaces. On the other hand, we want these wave functions
to live in the same space, such that we can form superpositions of wave functions
describing different numbers of particles. To accomplish this, we define
\begin{equation}\label{e3.14}
\Psi_n(x_{n,1}, \ldots , x_{n,n})=\sqrt{ \frac{{\cal V}^{(n)}}{{\cal V}} } \,
\psi_n(x_{n,1}, \ldots , x_{n,n}) ,
\end{equation}
where
\begin{equation}\label{e3.15}
 {\cal V}^{(n)}=\int d^4x_{n,1}\cdots \int d^4x_{n,n} ,
\end{equation}
\begin{equation}\label{e3.16}
 {\cal V}=\prod_{n=1}^{\infty} {\cal V}^{(n)} ,
\end{equation}
are volumes of the corresponding configuration spaces.
In particular, the wave function of the vacuum is
\begin{equation}\label{e3.17}
\Psi_0=\frac{1}{\sqrt{{\cal V}}} .
\end{equation}
This provides that all wave functions are normalized in the same configuration space
as
 \begin{equation}\label{e3.18}
\int {\cal D}\vec{x} \, | \Psi_n(x_{n,1}, \ldots , x_{n,n})|^2 =1 ,
\end{equation}
where we use the notation
\begin{equation}\label{e3.19}
 \vec{x}=(x_{1,1},x_{2,1},x_{2,2},\ldots ),
\end{equation}
\begin{equation}\label{e3.20}
 {\cal D}\vec{x} = \prod_{n=1}^{\infty} \, \prod_{a_{n}=1}^{n} d^4x_{n,a_{n}} .
\end{equation}

Note that the physical Hilbert space does not contain non-symmetrized
states, such as a 3-particle state $|x_{1,1}\rangle |x_{2,1},x_{2,2}\rangle$.
It also does not contain states that do not satisfy (\ref{e3.9'}). 
Nevertheless, the notation can be further simplified by introducing an extended
kinematic Hilbert space that contains such unphysical states as well.
Every physical state can be viewed as a state in such an extended Hilbert space,
although most of the states in the extended Hilbert space are not physical.
In this extended space it is convenient to denote the pair of
labels $(n,a_n)$ by a single label $A$. Hence,  (\ref{e3.19}) and (\ref{e3.20})
are now written as
\begin{equation}\label{e3.21}
 \vec{x}=(x_1,x_2,x_3,\ldots ),
\end{equation}
\begin{equation}\label{e3.22}
 {\cal D}\vec{x} = \prod_{A=1}^{\infty} d^4x_A .
\end{equation}
Similarly, (\ref{e3.16}) with (\ref{e3.15}) is now written as
\begin{equation}\label{e3.23}
 {\cal V}=\int \prod_{A=1}^{\infty} d^4x_A .
\end{equation}
The particle-position basis of this extended space is denoted by $|\vec{x})$ (which should be
distinguished from $|\vec{x}\rangle$ which would denote a symmetrized state
of an infinite number of physical particles).  
Such a basis allows us to write  
the physical wave function (\ref{e3.14}) as a wave function 
on the extended space 
\begin{equation}\label{e3.25}
 \Psi_n(\vec{x})=(\vec{x}|\Psi_n\rangle .
\end{equation}
Now (\ref{e3.18}) takes a simpler form 
\begin{equation}\label{e3.18'}
\int {\cal D}\vec{x} \, | \Psi_n(\vec{x})|^2 =1 .
\end{equation}
The normalization (\ref{e3.18'}) corresponds to the normalization 
in which the unit operator on the extended space is
\begin{equation}\label{e3.24}
 1=\int {\cal D}\vec{x} \, |\vec{x}) (\vec{x}| ,
\end{equation}
while the scalar product is
\begin{equation}\label{e3.24'}
 (\vec{x}|\vec{x}')=\delta(\vec{x}-\vec{x}') ,
\end{equation}
with $\delta(\vec{x}-\vec{x}') \equiv \prod_{A=1}^{\infty}\delta^4(x_A-x'_A)$.
A general physical state can be written as
\begin{equation}\label{e3.26}
 \Psi(\vec{x})=(\vec{x}|\Psi\rangle = 
\sum_{n=0}^{\infty}c_n \Psi_n(\vec{x}) .
\end{equation}
It is also convenient to write this as
\begin{equation}\label{e3.27}
\Psi(\vec{x})=\sum_{n=0}^{\infty} \tilde{\Psi}_n(\vec{x}) ,
\end{equation}
where the tilde denotes a wave function that is not necessarily normalized.
The total wave function is normalized, in the sense that
\begin{equation}\label{e3.26.1}
 \int {\cal D}\vec{x} \, |\Psi(\vec{x})|^2=1 ,
\end{equation}
implying
\begin{equation}\label{e3.26.2}
 \sum_{n=0}^{\infty}|c_n|^2=1 .
\end{equation}

Next, we introduce the operator
\begin{equation}\label{e3.28}
 \Box = \sum_{A=1}^{\infty} \partial_A^{\mu} \partial_{A\mu} .
\end{equation}
From the equations above (see, in particular, (\ref{e3.7})-(\ref{e3.12})), 
it is easy to show that $\Psi_n(\vec{x})$ satisfies
\begin{equation}\label{e3.29}
 \Box\Psi_n(\vec{x}) +nm^2 \Psi_n(\vec{x}) =0 .
\end{equation}
Introducing a hermitian number-operator $\hat{N}$ with the property
\begin{equation}\label{e3.30}
 \hat{N}\Psi_n(\vec{x}) = n \Psi_n(\vec{x}) ,
\end{equation}
one finds that a general physical state (\ref{e3.26}) satisfies the generalized
Klein-Gordon equation
\begin{equation}\label{e3.31}
 \Box\Psi(\vec{x})+m^2\hat{N}\Psi(\vec{x})=0.
\end{equation}
We also introduce the generalized Klein-Gordon current
\begin{equation}\label{e3.32}
 J^{\mu}_A(\vec{x})=
\frac{i}{2} \Psi^*(\vec{x})\!\stackrel{\leftrightarrow\;}{\partial^{\mu}_A}\! \Psi (\vec{x}) .
\end{equation}
From (\ref{e3.31}) one finds that, in general, this current is not conserved
\begin{equation}\label{e3.33}
 \sum_{A=1}^{\infty}\partial_{A\mu}J^{\mu}_A(\vec{x}) = J(\vec{x}) ,
\end{equation}
where
\begin{equation}\label{e3.34}
 J(\vec{x})=-\frac{i}{2}m^2 
\Psi^*(\vec{x})\!\stackrel{\;\leftrightarrow}{\hat{N}}\! \Psi (\vec{x}) ,
\end{equation}
and $\Psi' \!\stackrel{\;\leftrightarrow}{\hat{N}}\! \Psi \equiv 
\Psi' (\hat{N}\Psi) - ( \hat{N} \Psi') \Psi$. 
From (\ref{e3.34}) we see that the current is conserved in two special cases:
(i) when $\Psi=\Psi_n$ (a state with a definite number of physical
particles), or (ii) when $m^2=0$ (any physical state of massless particles). 

In the extended Hilbert space it is also useful to introduce the momentum picture
through the Fourier transforms. We define
\begin{equation}\label{e3.35}
 \Psi_{\vec{k}}(\vec{x})= \sqrt{ \frac{(2\pi)^{4\aleph_0}}{{\cal V}} } 
(\vec{x}|\vec{k})
=\frac{e^{-i\vec{k}\vec{x}}}{\sqrt{{\cal V}}} ,
\end{equation}
where $\vec{k}\vec{x}\equiv \sum_{A=1}^{\infty} k_{A\mu}x_A^{\mu}$
and $\aleph_0=\infty$ corresponds to the number of different values of the 
label $A$.
In the basis of momentum eigenstates $|\vec{k})$ we have
\begin{equation}\label{e3.36}
 1= \int {\cal D}\vec{k}\,  |\vec{k}) (\vec{k}| ,
\end{equation}
\begin{equation}\label{e3.37}
 (\vec{k}|\vec{k}')=\delta(\vec{k}-\vec{k}') .
\end{equation} 
It is easy to check that the normalizations as above make the Fourier transform
\begin{equation}\label{e3.38}
 \Psi(\vec{k})=(\vec{k}|\Psi\rangle=\int{\cal D}\vec{x}\, 
(\vec{k}|\vec{x})(\vec{x}|\Psi\rangle  
\end{equation}
and its inverse
\begin{equation}\label{e3.39}
 \Psi(\vec{x})=(\vec{x}|\Psi\rangle=\int {\cal D}\vec{k}\, 
(\vec{x}|\vec{k})(\vec{k}|\Psi\rangle 
\end{equation}
consistent. We can also introduce the momentum operator
\begin{equation}\label{e3.40}
\hat{p}_{A\mu}=i\partial_{A\mu}.
\end{equation}
The wave function (\ref{e3.35}) is the momentum eigenstate
\begin{equation}\label{e3.41}
 \hat{p}_{A\mu}\Psi_{\vec{k}}(\vec{x}) =
k_{A\mu}\Psi_{\vec{k}}(\vec{x}) .
\end{equation}
In particular, the wave function of the physical vacuum is given by (\ref{e3.17}), so
\begin{equation}\label{e3.42}
 \hat{p}_{A\mu}\Psi_0(\vec{x}) = 0 .
\end{equation}
We see that (\ref{e3.17}) can also be written as
\begin{equation}\label{e3.35'}
 \Psi_0(\vec{x})=\frac{e^{-i\vec{0}\vec{x}}}
{\sqrt{{\cal V}}} ,
\end{equation}
showing that the physical vacuum can also be represented as
\begin{equation}\label{e3.43}
 |0\rangle = |\vec{k}=\vec{0}) .
\end{equation}
Intuitively, it means that the vacuum can be thought of 
as a state with an infinite number of particles,
all of which have vanishing 4-momentum.
Similarly, an $n$-particle state can be thought of as a state with an infinite number of particles,
where only $n$ of them have a non-vanishing 4-momentum. 

Finally, let us rewrite some of the
main results of this (somewhat lengthy) subsection in a form
that will be suitable for a generalization in the next subsection.
A general physical state can be written in the form
\begin{equation}\label{s1}
|\Psi\rangle = \sum_{n=0}^{\infty} c_n |\Psi_n\rangle = 
\sum_{n=0}^{\infty}  |\tilde{\Psi}_n\rangle .
\end{equation} 
The corresponding unnormalized $n$-particle wave functions are
\begin{equation}\label{s2}
 \tilde{\psi}_n(x_{n,1},\ldots,x_{n,n}) =
\langle 0|\hat{\psi}_n(x_{n,1},\ldots,x_{n,n})|\Psi\rangle .
\end{equation}
There is a well-defined transformation 
\begin{equation}\label{s3}
 \tilde{\psi}_n(x_{n,1},\ldots,x_{n,n}) \rightarrow \tilde{\Psi}_n(\vec{x})  
\end{equation}
from the physical Hilbert space to the extended Hilbert space, so that
the general state (\ref{s1}) can be represented by a single wave function
\begin{equation}\label{s4}
\Psi(\vec{x})=\sum_{n=0}^{\infty} c_n \Psi_n(\vec{x})
=\sum_{n=0}^{\infty} \tilde{\Psi}_n(\vec{x}).
\end{equation}

\subsection{Generalization to the interacting QFT}
\label{SEC3.3}

In this subsection we discuss the generalization of the results of the preceding subsection
to the case in which the field operator $\hat{\phi}$ does not satisfy the free 
Klein-Gordon equation (\ref{e3.7}). For example,
if the classical action is
\begin{equation}\label{e3.44}
 S=\int d^4x  \left[ \frac{1}{2}(\partial^{\mu}\phi) (\partial_{\mu}\phi)
-\frac{m^2}{2}\phi^2 - \frac{\lambda}{4}\phi^4 \right] ,
\end{equation}
then (\ref{e3.7}) generalizes to
 \begin{equation}\label{e3.45}
 \partial^{\mu}\partial_{\mu}\hat{\phi}_H(x)+ 
m^2\hat{\phi}_H(x) +\lambda \hat{\phi}_H^3(x)=0, 
\end{equation}
where $\hat{\phi}_H(x)$ is the field operator in the Heisenberg picture.
(From this point of view, the operator $\hat{\phi}(x)$ 
defined by (\ref{e3.8}) and (\ref{e3.9}) and satisfying the free Klein-Gordon 
equation (\ref{e3.7}) is the field operator in the interaction (Dirac) picture.)
Thus, instead of (\ref{s2}) now we have
\begin{equation}\label{s2'}
 \tilde{\psi}_n(x_{n,1},\ldots,x_{n,n}) =
\langle 0|\hat{\psi}_{nH}(x_{n,1},\ldots,x_{n,n})|\Psi\rangle ,
\end{equation}
where $|\Psi\rangle$ and $|0\rangle$ are states in the Heisenberg picture.
Assuming that (\ref{s2'}) has been calculated (we shall see below how
in practice it can be done), the rest of the job is straightforward.
One needs to make the transformation (\ref{s3}) in the same way
as in the free case, which leads to an interacting variant of  (\ref{s4})
\begin{equation}\label{s4'}
\Psi(\vec{x})=\sum_{n=0}^{\infty} \tilde{\Psi}_n(\vec{x}) .
\end{equation}
The wave function (\ref{s4'}) encodes the complete information about the
properties of the interacting system. 

Now let us see how (\ref{s2'}) can be calculated in practice. Any operator
$\hat{O}_H(t)$ in the Heisenberg picture depending on a single time-variable $t$
can be written in terms of operators in the interaction picture as 
\begin{equation}\label{e3.46}
\hat{O}_H(t)=\hat{U}^{\dagger}(t)\hat{O}(t)\hat{U}(t) , 
\end{equation}
where
\begin{equation}\label{e3.47}
 \hat{U}(t)=Te^{-i\int_{t_0}^t dt' \hat{H}_{\rm int}(t')} ,
\end{equation}
$t_0$ is some appropriately chosen ``initial'' time, $T$ denotes the time ordering,
and $\hat{H}_{\rm int}$
is the interaction part of the Hamiltonian expressed as a functional of field operators
in the interaction picture (see, e.g., \cite{chengli}). 
For example, for the action (\ref{e3.44}) we have
\begin{equation}\label{e3.48}
 \hat{H}_{\rm int}(t)=\frac{\lambda}{4} \int d^3x \, :\!\hat{\phi}^4({\bf x},t)\!: ,
\end{equation}
where $:\;:$ denotes the normal ordering.
The relation (\ref{e3.46}) can also be inverted, leading to
\begin{equation}\label{e3.49}
\hat{O}(t)=\hat{U}(t)\hat{O}_H(t)\hat{U}^{\dagger}(t) . 
\end{equation}
Thus, the relation (\ref{e3.10}), which is now valid in the interaction picture,
allows us to write an analogous relation in the Heisenberg picture 
\begin{eqnarray}\label{e3.10'}
 \hat{\psi}_{nH}(x_{n,1}, \ldots , x_{n,n}) & = & d_n 
S_{ \{x_{n,1}, \ldots , x_{n,n} \} } 
\nonumber \\
& & \hat{\psi}_H(x_{n,1}) \cdots \hat{\psi}_H(x_{n,n}) ,
\end{eqnarray}
where
\begin{equation}\label{e3.50}
\hat{\psi}_H(x_{n,a_n})=\hat{U}^{\dagger}(x^0_{n,a_n})
\hat{\psi}(x_{n,a_n})\hat{U}(x^0_{n,a_n}) .
\end{equation}
By expanding (\ref{e3.47}) in powers of $\int_{t_0}^t dt' \hat{H}_{\rm int}$,
this allows us to calculate (\ref{e3.10'}) and (\ref{s2'}) perturbatively.
In (\ref{s2'}), the states in the Heisenberg picture $|\Psi\rangle$ and $|0\rangle$
are identified with the states in the interaction picture at the initial time
$|\Psi(t_0)\rangle$ and $|0(t_0)\rangle$, respectively.

To demonstrate that such a procedure leads to a physically sensible result,
let us see how it works in the special (and more familiar) case of the equal-time
wave function. It is given by $\tilde{\psi}_n(x_{n,1},\ldots,x_{n,n})$ 
calculated at $x^0_{n,1}=\cdots=x^0_{n,n}\equiv t$.
Thus, (\ref{s2'}) reduces to
\begin{eqnarray}\label{e3.51}
& \tilde{\psi}_n({\bf x}_{n,1},\ldots,{\bf x}_{n,n};t) = d_n  
\langle 0(t_0)|  \hat{U}^\dagger(t) \hat{\psi}({\bf x}_{n,1},t) \hat{U}(t) &
\nonumber \\
& \cdots
\hat{U}^\dagger(t) \hat{\psi}({\bf x}_{n,n},t) \hat{U}(t)   |\Psi(t_0)\rangle . & 
\end{eqnarray}
Using $\hat{U}(t)\hat{U}^\dagger(t)=1$ and
\begin{equation}\label{e3.52}
 \hat{U}(t)   |\Psi(t_0)\rangle = |\Psi(t)\rangle , \;\;\;\;
\hat{U}(t)   |0(t_0)\rangle = |0(t)\rangle ,
\end{equation}
the expression further simplifies 
\begin{eqnarray}\label{e3.53}
& \tilde{\psi}_n({\bf x}_{n,1},\ldots,{\bf x}_{n,n};t) = &
\nonumber \\
& d_n
\langle 0(t)|  \hat{\psi}({\bf x}_{n,1},t)  \cdots
\hat{\psi}({\bf x}_{n,n},t)    |\Psi(t)\rangle . &
\end{eqnarray}
In practical applications of QFT in particle physics, one usually calculates the 
$S$-matrix, corresponding to the limit $t_0\rightarrow -\infty$, 
$t\rightarrow\infty$. For Hamiltonians that conserve energy (such as (\ref{e3.48}))
this limit provides the stability of the vacuum, i.e., obeys
\begin{equation}\label{e3.54}
\lim_{t_0\rightarrow -\infty, \; t\rightarrow\infty} \hat{U}(t)   |0(t_0)\rangle = 
e^{-i\varphi_0} |0(t_0)\rangle , 
\end{equation}
where $\varphi_0$ is some physically irrelevant phase \cite{bd2}. 
Essentially, this is because the integrals of the type 
$\int_{-\infty}^{\infty} dt' \cdots$ produce $\delta$-functions
that correspond to energy conservation, so the vacuum remains stable
because particle creation from the vacuum would violate energy conservation.
Thus we have
\begin{equation}\label{e3.55}
 |0(\infty)\rangle =e^{-i\varphi_0}|0(-\infty)\rangle \equiv  e^{-i\varphi_0}|0\rangle .
\end{equation}
The state
\begin{equation}\label{e3.55.1}
 |\Psi(\infty)\rangle=\hat{U}(\infty)|\Psi(-\infty)\rangle 
\end{equation}
is not trivial, but whatever it is, it has some expansion of the form
\begin{equation}\label{e3.56}
 |\Psi(\infty)\rangle =  \sum_{n=0}^{\infty} c_{n}(\infty)|\Psi_{n}\rangle , 
\end{equation}
where $c_{n}(\infty)$ are some coefficients. 
Plugging (\ref{e3.55}) and (\ref{e3.56})
into (\ref{e3.53}) and recalling (\ref{e3.10})-(\ref{e3.12.1}),
we finally obtain
\begin{equation}\label{e3.57}
 \tilde{\psi}_n({\bf x}_{n,1},\ldots,{\bf x}_{n,n};\infty)=
e^{i\varphi_0} c_n(\infty) \psi_n({\bf x}_{n,1},\ldots,{\bf x}_{n,n};\infty) .
\end{equation}
This demonstrates the consistency of (\ref{s2'}), because (\ref{e3.55.1}) 
should be recognized as the standard description of evolution from
$t_0\rightarrow -\infty$ to $t\rightarrow \infty$ (see, e.g., \cite{chengli,bd2}),
showing that the coefficients $c_n(\infty)$ are the same as those described 
by standard $S$-matrix theory in QFT.
In other words, (\ref{s2'}) is a natural many-time generalization of the concept of 
single-time evolution in interacting QFT.

\subsection{Generalization to other types of particles}
\label{SEC3.4}

In Secs.~\ref{SEC3.2} and \ref{SEC3.3} we have discussed in detail
scalar hermitian fields, corresponding to spinless uncharged particles.
In this subsection we briefly discuss how these results can be
generalized to any type of fields and the corresponding particles.

In general, fields $\phi$ carry some additional labels which we 
collectively denote by $l$, so we deal with fields $\phi_l$.
For example, spin-1 field carries a
vector index, fermionic spin-$\frac{1}{2}$ field carries a spinor index,
non-Abelian gauge fields carry internal indices of the gauge group, etc.
Thus Eq.~(\ref{e3.10}) generalizes to
\begin{eqnarray}\label{e3.10gen}
 & \hat{\psi}_{n,L_n}(x_{n,1}, \ldots , x_{n,n}) = & 
\nonumber \\
& d_n 
S_{ \{x_{n,1}, \ldots , x_{n,n} \} } 
\hat{\psi}_{l_{n,1}}(x_{n,1}) \cdots \hat{\psi}_{l_{n,n}} (x_{n,n}) , &
\end{eqnarray}
where $L_n$ is a collective label $L_n=(l_{n,1}, \ldots , l_{n,n} )$.
The symbol $S_{ \{x_{n,1}, \ldots , x_{n,n} \} }$ denotes symmetrization (antisymmetrization) over bosonic (fermionic) fields describing the same type of particles.
Hence, it is straightforward to make the appropriate generalizations of all results
of Secs.~\ref{SEC3.2} and \ref{SEC3.3}. For example, (\ref{e3.27})
generalizes to 
\begin{equation}\label{e3.27gen}
\Psi_{\vec{L}} (\vec{x}) = \sum_{n=0}^{\infty} \sum_{L_n} \tilde{\Psi}_{n,L_n}
(\vec{x}) ,
\end{equation}
with self-explaining notation.

To further simplify the notation, we introduce the column 
$\Psi\equiv \{\Psi_{\vec{L}} \}$ and the row
$\Psi^{\dagger}\equiv \{\Psi^*_{\vec{L}} \}$.
With this notation, the appropriate generalization of (\ref{e3.26.1}) can be written as
\begin{equation}\label{e3.26.1gen}
 \int {\cal D}\vec{x} \, \sum_{{\vec{L}}} \Psi^*_{\vec{L}}(\vec{x})
\Psi_{\vec{L}}(\vec{x}) \equiv \int {\cal D}\vec{x} \,
\Psi^{\dagger}(\vec{x}) \Psi(\vec{x})
=1 .
\end{equation}

For the case of states that contain fermionic particles, Eq.~(\ref{e3.26.1gen})
requires further discussion. As a simple example, consider a 
1-particle state describing one electron.
In this case, (\ref{e3.26.1gen}) can be reduced to
\begin{equation}\label{e3.58}
  \int d^4x \, \psi^{\dagger}(x)\psi(x) =1 ,
\end{equation}
where $\psi$ is a Dirac spinor. In this expression, the quantity $\psi^{\dagger}\psi$
must transform as a Lorentz scalar. At first sight, it may seem to be in contradiction
with the well-known fact that $\psi^{\dagger}\psi=\bar{\psi}\gamma^0\psi$
transforms as a time-component of a Lorentz vector. However, there is no 
true contradiction. Let us explain.

The standard derivation that $\bar{\psi}\gamma^{\mu}\psi$ transforms as a vector \cite{bd1}
starts from the assumption that the matrices $\gamma^{\mu}$  do not transform
under Lorentz transformations, despite of carrying the index ${\mu}$.
However, such an assumption
is not necessary. Moreover, in curved spacetime such an assumption is inconsistent \cite{bd}.
In fact, one is allowed to define the transformations of $\psi$ and $\gamma^{\mu}$
in an arbitrary way, as long as such transformations do not affect the transformations
of measurable quantities, or quantities like $\bar{\psi}\gamma^{\mu}\psi$ 
that are closely related to measurable ones. 
Thus, it is much more natural to deal with a differently defined transformations
of $\gamma^{\mu}$ and $\psi$, such that $\gamma^{\mu}$ transforms as a vector
and $\psi$ transforms as a scalar under Lorentz transformations of spacetime coordinates \cite{bd}. 
The spinor indices of $\gamma^{\mu}$ and $\psi$ are then reinterpreted as indices
in an internal space. With such redefined transformations, (\ref {e3.58}) is fully consistent.
The details of our transformation conventions are presented in Appendix \ref{APPA}.

\section{The physical interpretation}
\label{SEC4}

\subsection{Probabilistic interpretation}

In this subsection we adopt and further develop the probabilistic interpretation
introduced in \cite{nikprobrel09} (and partially inspired by earlier results 
\cite{stuc1,stuc2,horw,kypr,fanchi}).
The quantity
\begin{equation}\label{e4.1}
 {\cal D}P=\Psi^{\dagger}(\vec{x}) \Psi(\vec{x}) \, {\cal D}\vec{x}
\end{equation}
is naturally interpreted as the probability of finding the system in the 
(infinitesimal) configuration-space volume ${\cal D}\vec{x}$ around a 
point $\vec{x}$ in the configuration space. Indeed, such an interpretation
is consistent with our normalization conditions such as 
(\ref{e3.26.1}) and (\ref{e3.26.1gen}). In more physical terms, it gives the 
joint probability that the particle $1$ is found at the spacetime position 
$x_1$,  particle $2$ at the spacetime position $x_2$, etc.
Similarly, the Fourier-transformed
wave function $\Psi(\vec{k})$ defines the probability 
$\Psi^{\dagger}(\vec{k}) \Psi(\vec{k}) {\cal D}\vec{k}$, which is the 
joint probability that the particle $1$ has the 4-momentum
$k_1$,  particle $2$ the 4-momentum $k_2$, etc.

As a special case, consider an $n$-particle state
$\Psi(\vec{x})=\Psi_n(\vec{x})$. It really depends only on
$n$ spacetime positions $x_{n,1},\ldots x_{n,n}$. With respect to
all other positions $x_B$, $\Psi$ is a constant. Thus, the probability
of various positions  $x_B$ does not depend on $x_B$; such a particle can be found
anywhere and anytime with equal probabilities. There is an infinite number of such
particles. Nevertheless, the Fourier transform of such a wave function reveals
that the 4-momentum $k_B$ of these particles is necessarily zero; they have neither
3-momentum nor energy. For that reason, such particles can be thought of as ``vacuum'' particles. In this picture, an $n$-particle state $\Psi_n$ is thought of as a state describing
$n$ ``real'' particles and an infinite number of ``vacuum'' particles.

To avoid a possible confusion with the usual notions of vacuum  
and real particles in QFT, in the rest of the paper
we refer to ``vacuum'' particles as {\it dead} particles
and ``real'' particles as {\it live} particles. Or let us be 
slightly more precise: We say that the 
particle $A$ is dead if the wave function in the momentum space
$\Psi(\vec{k})$ vanishes for all values of $k_A$ except $k_A=0$.
Similarly, we say that the particle $A$ is live if it is not dead.

The properties of live particles associated with the state $\Psi_n(\vec{x})$ can also be 
represented by the wave function $\psi_n(x_{n,1},\ldots,x_{n,n} )$. By averaging over
physically uninteresting dead particles, (\ref{e4.1}) reduces to
 \begin{eqnarray}\label{e4.2}
 dP & = & \psi_n^{\dagger}(x_{n,1},\ldots,x_{n,n}) \psi_n(x_{n,1},\ldots,x_{n,n})
\nonumber \\ 
& & \times \, d^4x_{n,1}\cdots d^4x_{n,n},
\end{eqnarray}
which involves only live particles. This describes the probability when neither the space
positions of detected particles nor times of their detections are known.
To relate it with a more familiar probabilistic interpretation of QM, let us consider
the special case; let us assume that the first particle 
is detected at time $x^0_{n,1}$, second particle at time $x^0_{n,2}$, etc.
In this case, the detection times are known, so (\ref{e4.2}) is no longer
the best description of our knowledge about the system. Instead, the relevant 
probability derived from (\ref{e4.2})
is the {\em conditional} probability
\begin{eqnarray}\label{e4.3}
 dP_{(3n)} & = & 
\frac{ \psi_n^{\dagger}(x_{n,1},\ldots,x_{n,n}) \psi_n(x_{n,1},\ldots,x_{n,n}) }
{N_{x^0_{n,1},\cdots,x^0_{n,n}}}
\nonumber \\ 
& & \times \, d^3x_{n,1}\cdots d^3x_{n,n},
\end{eqnarray}
where
\begin{eqnarray}\label{e4.4}
 N_{x^0_{n,1},\cdots,x^0_{n,n}} & = & \int \psi_n^{\dagger}(x_{n,1},\ldots,x_{n,n})
\psi_n(x_{n,1},\ldots,x_{n,n}) 
\nonumber \\ 
& & \times \, d^3x_{n,1}\cdots d^3x_{n,n}
\end{eqnarray}
is the appropriate normalization factor. 
(For more details regarding the meaning and limitations of (\ref{e4.3}) in the 
1-particle case see Appendix \ref{APPB}.)
The probability (\ref{e4.3}) is sometimes also postulated as a fundamental axiom
of many-time formulation of QM \cite{tomonaga}, but here  
(\ref{e4.3}) is derived from a more fundamental and more general expression (\ref{e4.2}) (which, in turn, is derived from an even more general axiom (\ref{e4.1})). 
An even more familiar expression is obtained by studding a special case
of (\ref{e4.3}) in which $x^0_{n,1}=\cdots =x^0_{n,n}\equiv t$, so that
(\ref{e4.3}) reduces to
\begin{eqnarray}\label{e4.5}
 dP_{(3n)} & = & 
\frac{ \psi_n^{\dagger}({\bf x}_{n,1},\ldots,{\bf x}_{n,n};t) 
\psi_n({\bf x}_{n,1},\ldots,{\bf x}_{n,n};t) }
{N_{t}}
\nonumber \\ 
& & \times \, d^3x_{n,1}\cdots d^3x_{n,n},
\end{eqnarray}
where $N_{t}$ is given by (\ref{e4.4}) evaluated at $x^0_{n,1}=\cdots =x^0_{n,n}\equiv t$.

Now let us see how the wave functions representing the states in interacting QFT are interpreted
probabilistically. Consider the wave function $\tilde{\psi}_n(x_{n,1},\ldots,x_{n,n})$
given by (\ref{s2'}). For example, it may vanish for small values of 
$x^0_{n,1}, \dots, x^0_{n,n}$, but it may not vanish for their large values. Physically, it means
that these particles cannot be detected in the far past (the probability is zero), but that
they can be detected in the far future. This is nothing but a probabilistic description of
the creation of $n$ particles that have not existed in the far past. Indeed, 
the results obtained in Sec.~\ref{SEC3.3} (see, in particular, (\ref{e3.57}))
show that such probabilities are consistent with the probabilities of particle creation obtained
by the standard $S$-matrix methods in QFT.

Having developed the probabilistic interpretation, we can also calculate the average values
of various quantities. We are particularly interested in average values of the 4-momentum
$p^{\mu}_A$. In general, its average value is
\begin{equation}\label{e4.6}
 \langle p^{\mu}_A \rangle= \int {\cal D}\vec{x} \, \Psi^{\dagger}(\vec{x}) \hat{p}^{\mu}_A
\Psi(\vec{x}) ,
\end{equation}
where $\hat{p}^{\mu}_A$ is given by (\ref{e3.40}). 
If $\Psi(\vec{x})=\Psi_n(\vec{x})$, then (\ref{e4.6}) can be reduced to
\begin{eqnarray}\label{e4.7}
 \langle p^{\mu}_{n,a_n} \rangle & = & \int  d^4x_{n,1}\cdots d^4x_{n,n}
\\
& & \psi_n^{\dagger}(x_{n,1},\ldots,x_{n,n}) \hat{p}^{\mu}_{n,a_n}
\psi_n(x_{n,1},\ldots,x_{n,n}) .
\nonumber
\end{eqnarray}
Similarly, if the times of detections are known and are all equal to $t$, then the average
space-components of momenta are given by a more familiar expression
\begin{eqnarray}\label{e4.8}
 \langle {\bf p}_{n,a_n} \rangle & = & N_t^{-1} \int  d^3x_{n,1}\cdots d^3x_{n,n}
\\
& & \psi_n^{\dagger}({\bf x}_{n,1},\ldots,{\bf x}_{n,n};t) \hat{{\bf p}}_{n,a_n}
\psi_n({\bf x}_{n,1},\ldots,{\bf x}_{n,n};t) .
\nonumber
\end{eqnarray}

Finally, note that (\ref{e4.6}) can also be written in an alternative form 
\begin{equation}\label{e4.9}
 \langle p^{\mu}_A \rangle= \int {\cal D}\vec{x}\, \rho(\vec{x}) U^{\mu}_A(\vec{x}) ,
\end{equation}
where
\begin{equation}\label{e4.10}
\rho(\vec{x})=\Psi^{\dagger}(\vec{x})\Psi(\vec{x})
\end{equation}
is the probability density and
\begin{equation}\label{e4.11}
 U^{\mu}_A(\vec{x})=\frac{J^{\mu}_A(\vec{x})}{\Psi^{\dagger}(\vec{x})\Psi(\vec{x})} .
\end{equation}
Here $J^{\mu}_A$ is given by an obvious generalization of (\ref{e3.32}) 
\begin{equation}\label{e4.12}
 J^{\mu}_A(\vec{x})=
\frac{i}{2} \Psi^{\dagger}(\vec{x})\!\stackrel{\leftrightarrow\;}{\partial^{\mu}_A}\! 
\Psi (\vec{x}) .
\end{equation}
The expression (\ref{e4.9}) will play an important role in the next subsection.

\subsection{Particle-trajectory interpretation}

The idea of the particle-trajectory interpretation is that each particle has some trajectory
$X^{\mu}_A(s)$, where $s$ is an auxiliary scalar parameter that parameterizes the trajectories.
Such trajectories must be consistent with the probabilistic interpretation (\ref{e4.1}).
Thus, we need a velocity function $V^{\mu}_A(\vec{x})$, so that the trajectories
satisfy
\begin{equation}\label{e4.13}
 \frac{dX^{\mu}_A(s)}{ds}=V^{\mu}_A(\vec{X}(s)) ,
\end{equation}
where the velocity function must be such that the following conservation equation is obeyed
\begin{equation}\label{e4.14}
 \frac{\partial \rho(\vec{x})}{\partial s} + 
\sum_{A=1}^{\infty}\partial_{A\mu}[\rho(\vec{x}) V^{\mu}_A(\vec{x}) ] =0.
\end{equation}
Namely, if a statistical ensemble of particle positions in spacetime has the distribution 
(\ref{e4.10}) for some initial $s$, then (\ref{e4.13}) and (\ref{e4.14}) will 
provide that this statistical ensemble will also have the distribution 
(\ref{e4.10}) for {\em any} $s$, making the trajectories consistent with (\ref{e4.1}).
The first term in (\ref{e4.14}) trivially vanishes: $ \partial \rho(\vec{x})/\partial s =0$.
Thus, the condition (\ref{e4.14}) reduces to the requirement
\begin{equation}\label{e4.15} 
\sum_{A=1}^{\infty}\partial_{A\mu}[\rho(\vec{x}) V^{\mu}_A(\vec{x}) ] =0.
\end{equation}
In addition, we require that the average velocity should be proportional to the average momentum
(\ref{e4.9}), i.e., 
\begin{equation}\label{e4.16}
\int {\cal D}\vec{x}\, \rho(\vec{x}) V^{\mu}_A(\vec{x}) 
 = {\rm const} \times \int {\cal D}\vec{x}\, \rho(\vec{x}) U^{\mu}_A(\vec{x}) .
\end{equation} 
In fact, the constant in (\ref{e4.16}) is physically irrelevant, because it can allways be
absorbed into a rescaling of the parameter $s$ in (\ref{e4.13}). The physical 3-velocity
$dX^{i}_A/dX^{0}_A$, $i=1,2,3$, is not affected by such a rescaling. Thus, in the rest
of the analysis we fix
\begin{equation}\label{e4.17}
{\rm const}=1 .
\end{equation}

As a first guess, Eq.~(\ref{e4.16}) with (\ref{e4.17}) suggests that one could take
$V^{\mu}_A=U^{\mu}_A$. However, it does not work in general. Namely, 
from (\ref{e4.10}) and (\ref{e4.11}) we see that 
$\rho U^{\mu}_A =J^{\mu}_A$, and we have seen in (\ref{e3.33})
that $ J^{\mu}_A$ does not need to be conserved. Instead, we have
\begin{equation}\label{e4.18}
 \sum_{A=1}^{\infty}\partial_{A\mu}[\rho(\vec{x}) U^{\mu}_A(\vec{x})] = J(\vec{x}) ,
\end{equation}
where $J(\vec{x})$ is some function that can be calculated explicitly whenever 
$\Psi(\vec{x})$ is known. So, how to find the appropriate function $V^{\mu}_A(\vec{x})$?

The problem of finding $V^{\mu}_A$ is solved in \cite{nikcr2} for a very general case
(see also \cite{struy}). Since the detailed derivation is presented in \cite{nikcr2},
here we only present the final results. Applying the general method developed in \cite{nikcr2},
one obtains
\begin{equation}\label{e4.19}
 V^{\mu}_A(\vec{x})=U^{\mu}_A(\vec{x}) +
\rho^{-1}(\vec{x}) [e^{\mu}_A + E^{\mu}_A(\vec{x})] ,
\end{equation}
where
\begin{equation}\label{e4.20}
 e^{\mu}_A=-{\cal V}^{-1} \int {\cal D}\vec{x}\, E^{\mu}_A(\vec{x}) ,
\end{equation}
\begin{equation}\label{e4.21}
 E^{\mu}_A(\vec{x})=\partial^{\mu}_A 
\int {\cal D}\vec{x}'\, G(\vec{x},\vec{x}') J(\vec{x}') ,
\end{equation}
\begin{equation}\label{e4.22}
 G(\vec{x},\vec{x}') = \int \frac{{\cal D}\vec{k}}{(2\pi)^{4\aleph_0}} 
\frac{e^{i\vec{k}(\vec{x}-\vec{x}')}}{\vec{k}^2} .
\end{equation}
Eqs.~(\ref{e4.21})-(\ref{e4.22}) provide that (\ref{e4.19}) obeys (\ref{e4.15}), 
while (\ref{e4.20}) provides that (\ref{e4.19}) obeys (\ref{e4.16})-(\ref{e4.17}).

We note two important properties of (\ref{e4.19}). First, if $J=0$ in (\ref{e4.18}),
then $V^{\mu}_A=U^{\mu}_A$. In particular, since $J=0$ for free fields in states
with a definite number of particles (it can be derived for any type of particles 
analogously to the derivation of (\ref{e3.34}) for spinless uncharged particles), 
it follows that $V^{\mu}_A=U^{\mu}_A$ for such states.
Second, if $\Psi(\vec{x})$ does not depend
on some coordinate $x^{\mu}_B$, then both $U^{\mu}_B=0$ and
$V^{\mu}_B=0$. [To show that $V^{\mu}_B=0$, note first that $J(\vec{x})$ defined
by (\ref{e4.18}) does not depend on $x^{\mu}_B$ when $\Psi(\vec{x})$ does not depend
on $x^{\mu}_B$. Then the integration over $dx'^{\mu}_B$ in (\ref{e4.21}) produces
$\delta(k^{\mu}_B)$, which kills the dependence on $x^{\mu}_B$ carried by 
(\ref{e4.22})].
This implies that dead particles have zero 4-velocity.

The results above show that the relativistic Bohmian trajectories are compatible with the
spacetime probabilistic interpretation (\ref{e4.1}). But what about the more
conventional space probabilistic interpretations (\ref{e4.3}) and (\ref{e4.5})?
{\it A priori}, these Bohmian trajectories are not compatible with (\ref{e4.3}) and (\ref{e4.5}).
Nevertheless, as discussed in more detail in Appendix \ref{APPB} for the 1-particle case,
the compatibility between measurable predictions of the Bohmian interpretation and that of the
``standard'' purely probabilistic interpretation restores when the appropriate theory
of quantum measurements is also taken into account.

Having established the general theory of particle trajectories by the results above, 
now we can discuss particular consequences. 

The trajectories are determined uniquely if the initial spacetime positions
$X^{\mu}_A(0)$ in (\ref{e4.13}), for all $\mu=0,1,2,3$, $A=1,\ldots, \infty$,
are specified. In particular,
since dead particles have zero 4-velocity, such particles do not really have trajectories
in spacetime. Instead, they are represented by dots in spacetime, as in 
Fig.~\ref{fig2} (Sec.~\ref{SEC2}). The spacetime positions of these dots are 
specified by their initial spacetime positions.

Since $\rho(\vec{x})$ describes probabilities for particle creation and destruction, 
and since (\ref{e4.14}) provides that particle trajectories are such that 
spacetime positions of particles
are distributed according to $\rho(\vec{x})$, it implies that particle trajectories are
also consistent with particle creation and destruction. In particular, the trajectories
in spacetime may have beginning and ending points, which correspond to points
at which their 4-velocities vanish (for an example, see Fig.~\ref{fig1}). For example,
the 4-velocity of the particle A vanishes if the conditional wave function
$\Psi(x_A,\vec{X}')$ does not depend on $x_A$ (where $\vec{X}'$ denotes
the actual spacetime positions of all particles except the particle $A$).

One very efficient mechanism of destroying particles is through the interaction
with the environment, such that the total quantum state takes the form 
(\ref{e3.4}). The environment wave functions
$(\vec{x}|E_0\rangle$, $(\vec{x}|E_{n;i_1,\ldots,i_n} \rangle$ do not overlap,
so the particles describing the environment can be in the support of only one of these
environment wave functions. Consequently, the conditional wave function
is described by only one of the terms in the sum (\ref{e3.4}), which effectively
collapses the wave function to only one of the terms in (\ref{e3.3}). For example,
if the latter wave function is $(\vec{x}|i_1,\ldots,i_n \rangle$, then it depends on only
$n$ coordinates among all $x_A$. All other live particles from sectors with
$n'\neq n$ become dead, i.e., their 4-velocities become zero which appears as
their destruction in spacetime.
More generally, if the overlap between the environment wave functions is negligible 
but not exactly zero, then particles from sectors with
$n'\neq n$ will not become dead, but their influence on the environment will still be
negligible, which still provides an effective collapse to $(\vec{x}|i_1,\ldots,i_n \rangle$.
Since decoherence is practically irreversible (due to many degrees of freedom involved), 
such an effective collapse is irreversible as well.

Another physically interesting situation is when the entanglement with the environment
takes the form (\ref{e3.6}), where $|\xi\rangle$ are coherent states. 
In this case, the behavior of the environment can very well be described in terms 
of an environment that responds to a presence of classical fields.
This explains how classical fields may appear at the
macroscopic level, even when the microscopic ontology is described
in terms of particles. 
Since $|\xi\rangle$ is a superposition of states with all possible numbers of particles,
trajectories of particles from sectors with different numbers of particles coexist;
there is an infinite number of live particle trajectories in that case. 

Similarly, an entanglement of the form of (\ref{u3.4}) explains how accelerated detectors
and detectors in a gravitational field may detect particles in the vacuum. 
For example, let us consider the case of a uniformly accelerated detector. In this case,
$|0'\rangle$ corresponds to the Rindler vacuum, while $|0\rangle$  is referred to as
the Minkowski vacuum \cite{bd}. The particle trajectories described by (\ref{e4.13})
are those of the Minkowski particles.
The interaction between the Minkowski vacuum and the accelerated detector creates 
new Minkowski particles.
For instance, if the detector is found in the state $|E_0\rangle$, then the Minkowski
particles are in the state $|0'\rangle$, which is a squeezed state 
describing an infinite number of live particle trajectories. 
Such a view seems particularly appealing from the point of view of
recently discovered renormalizable Horava-Lifshitz gravity \cite{horava}
that contains an absolute time and thus a preferred definition of particles
in a classical gravitational background \cite{nikhorava}. 

Let us also give a few remarks on measurements of 4-momenta and 4-velocities.
If $|i\rangle$ in (\ref{e3.1}) denote the 4-momentum eigenstates, then 
(\ref{e3.4}) describes a measurement of the particle 4-momenta. Since the 4-momentum 
eigenstates are also the 4-velocity eigenstates, (\ref{e3.4}) also describes a measurement 
of the particle 4-velocities. Thus, as discussed also in more detail in \cite{nikcr1}, even though
the Bohmian particle velocities may exceed the velocity of light, they cannot exceed 
the velocity of light when their velocities are measured. Instead, the effective wave function associated
with such a measurement is a momentum eigenstate of the form of (\ref{e3.35}),
where $k_A^2=m^2$ for live particles and  $k_A=0$ for dead particles.
This also explains why dead particles are not seen in experiments: their 3-momenta 
and energies are equal to zero.

Finally, we want to end this subsection with some conceptual remarks concerning
the physical meaning of the parameter $s$.
This parameter can be thought of as an evolution parameter, playing 
a role similar to that of the absolute Newton time $t$ in the usual formulation
of nonrelativistic Bohmian interpretation \cite{bohm1,bohm1}. To make the similarity
with such a usual formulation more explicit, it may be useful to think of $s$
as a coordinate parameterizing a ``fifth dimension'' that exists independently 
of other 4 dimensions with coordinates $x^{\mu}$. However, such a 
5-dimensional view should not be taken too literally. 
In particular, while the time $t$ is 
measurable, the parameter $s$ is not measurable.  

Given the fact that $s$ is not measurable, what is the physical meaning 
of the claim that particles have the distribution $\rho({\vec x})$ at some $s$?
We can think of it in the following way: We allways measure spacetime positions
of particles at some values of $s$, but we do not know what these values are.
Consequently, the probability density that describes our knowledge is described
by $\rho({\vec x})$ averaged over all possible values of $s$. However, since
$\rho({\vec x})$ does not depend on $s$, the result of such an averaging 
procedure is trivial, giving $\rho({\vec x})$ itself.

\section{Conclusion}
\label{SEC5}

In this paper we have extended the Bohmian interpretation of QM, such that
it also incorporates a description of particle creation and destruction described
by QFT. Unlike the previous attempts 
\cite{durrcr1,durrcr2} and \cite{nikcr1,nikcr2} to describe the creation and destruction
of pointlike particles within the Bohmian interpretation, the approach
of the present paper incorporates the creation and destruction
of pointlike particles {\em automatically}, without adding any additional
structure not already present in the equations that describe the continuous 
particle trajectories.
One reason why it works is the fact that we work with a many-time
wave function, so that the 4-velocity of each particle may vanish separately.
Even though the many-time wave function plays a central role,
we emphasize that the many-time wave function, 
first introduced in \cite{tomonaga}, is a natural concept
when one wants to treat time on an equal footing with space, even if one does
not have an ambition to describe particle creation and destruction 
\cite{nikprobrel09}.
Another, even more important reason why it works is the entanglement with the
environment, which explains an effective wave function collapse into
particle-number eigenstates (or some other eigenstates)
even when particles are not really created or destroyed.

As a byproduct, in this paper we have also obtained many technical results that allow to represent
QFT states with uncertain number of particles 
in terms of many-time wave functions. These results may be 
useful by they own, even without the Bohmian interpretation (see, e.g., \cite{nikpure}).

\section*{Acknowledgements}

This work was supported by the Ministry of Science of the
Republic of Croatia under Contract No.~098-0982930-2864.

\appendix

\section{Spinors and coordinate transformations}
\label{APPA}

At each point of spacetime, one can introduce the tetrad $e^{\mu}_{\bar{\alpha}}(x)$, which
is a collection of four spacetime vectors, $\bar{\alpha}=0,1,2,3$. 
The bar on $\bar{\alpha}$ denotes
that $\bar{\alpha}$ is not a spacetime-vector index, but only a label. By contrast, 
the index $\mu$ is a spacetime-vector index. 
The tetrad is chosen so that
\begin{equation}\label{a1}
\eta^{\bar{\alpha}\bar{\beta}}e^{\mu}_{\bar{\alpha}}(x)e^{\nu}_{\bar{\beta}}(x)=
g^{\mu\nu}(x) ,
\end{equation}
where $g^{\mu\nu}(x)$ is the spacetime metric and $\eta_{\bar{\alpha}\bar{\beta}}$ are 
components of a matrix equal to the Minkowski metric. 
The spacetime-vector indices are raised and lowered by $g^{\mu\nu}(x)$ and $g_{\mu\nu}(x)$,
respectively, while $\bar{\alpha}$-labels are raised and lowered by $\eta^{\bar{\alpha}\bar{\beta}}$ 
and $\eta_{\bar{\alpha}\bar{\beta}}$, respectively. Thus, (\ref{a1}) can also be inverted as
\begin{equation}\label{a2}
g^{\mu\nu}(x)e_{\mu}^{\bar{\alpha}}(x)e_{\nu}^{\bar{\beta}}(x)=
\eta^{\bar{\alpha}\bar{\beta}} .
\end{equation} 

Now let $\gamma^{\bar{\alpha}}$ be the standard Dirac matrices \cite{bd1}. From them we define
\begin{equation}\label{a3}
\gamma^{\mu}(x)=e^{\mu}_{\bar{\alpha}}(x) \gamma^{\bar{\alpha}} .
\end{equation}   
The spinor indices carried by matrices $\gamma^{\bar{\alpha}}$ and $\gamma^{\mu}(x)$
are interpreted as indices of the spinor representation of the {\em internal} 
group SO(1,3). Thus, $\gamma^{\bar{\alpha}}$ transform as scalars under spacetime coordinate 
transformations. Similarly, the spinors $\psi(x)$ are also scalars with respect to
spacetime coordinate transformations, while their spinor indices are indices in the
internal group SO(1,3). Likewise, $\psi^{\dagger}(x)$ is also a scalar with respect to
spacetime coordinate transformations. It is also convenient to define the quantity
\begin{equation}\label{a4}
\bar{\psi}(x)=\psi^{\dagger}(x) \gamma^{\bar{0}} ,
\end{equation}
which is also a scalar with respect to spacetime coordinate transformations.
Thus we see that the quantities 
\begin{equation}\label{a5}
\bar{\psi}(x)\psi(x) , \;\;\;\; \psi^{\dagger}(x)\psi(x) ,
\end{equation}
are both scalars with respect to spacetime coordinate transformations, and that
the quantities 
\begin{equation}\label{a6}
\bar{\psi}(x)\gamma^{\mu}(x)\psi(x) , \;\;\;\; 
i\psi^{\dagger}(x)  \!\stackrel{\leftrightarrow\;}{\partial^{\mu}}\!  \psi(x) ,
\end{equation}
are both vectors with respect to spacetime coordinate transformations.

Finally, we note that the relation with the more familiar (but less general) 
formalism in Minkowski spacetime
\cite{bd1} can be established by using the fact that in flat spacetime 
there is a global choice of coordinates in which
\begin{equation}\label{a7}
\gamma^{\mu}(x)=\gamma^{\bar{\mu}} .
\end{equation} 
However, (\ref{a7}) is not a covariant expression, but is only valid in one
special system of coordinates.

\section{Spacetime probability density and space probability density}
\label{APPB}

In this Appendix we present a more carefull discussion
of the meaning of our probabilistic interpretation that treats time on an equal
footing with space.
The first subsection deals 
with the pure probabilistic interpretation that does not assume the existence
of particle trajectories, while the second subsection deals with the
Bohmian interpretation in which the existence of particle trajectories is
also taken into account. For simplicity, we study the case of 1 particle only,
while the generalization to many particles is straightforward.

\subsection{Pure probabilistic interpretation}
\label{secAPPB1}

In a pure probabilistic interpretation of QM, one does not assume the existence
of particle trajectories. Instead, the main axiom is that $\psi(x)$ determines the
spacetime probability density
\begin{equation}\label{appb1}
 dP=|\psi(x)|^2 d^4x .
\end{equation}
This probability describes a statistical ensemble consisting of a large 
number of events, where each event is an appearance of the particle at the
spacetime point $x$. Now, if we pick up a subensemble consisting of all
events $x$ that have the same time coordinate $x^0=t$, then the distribution
in this subensemble is given by the conditional probability
\begin{equation}\label{appb2}
 dP_{(3)}=\frac{|\psi({\bf x},t)|^2 d^3x}{N_t} ,
\end{equation}
where the normalization factor $N_t=\int d^3x |\psi({\bf x},t)|^2$
is equal to the marginal probability that the particle from the
initial ensemble will have $x^0=t$. (In fact, when $\psi(x)$ is a
superposition of plane waves with positive frequencies only, then 
$N_t$ does not depend on $t$ \cite{nikprob}.)

A more formal way to understand (\ref{appb1}) is as follows.
We start from the identity
\begin{equation}\label{appb3}
\psi(x)=\int d^4x' \, \psi(x') \delta^4(x'-x) .
\end{equation}
Formally, it is convenient to use a discrete notation
\begin{equation}\label{appb4}
 \int d^4x' \rightarrow \sum_{x'} , \;\;\;\;
\psi(x') \rightarrow c_{x'} , \;\;\;\;
\delta^4(x'-x)  \rightarrow \psi_{x'}(x),
\end{equation}
where the function $\psi_{x'}(x)$ can be thought of as an eigenstate of the
operator $x$ with the eigenvalue $x'$. Thus, (\ref{appb3}) can be written as
\begin{equation}\label{appb5}
\psi(x)=\sum_{x'} c_{x'} \psi_{x'}(x) .
\end{equation}
The probability that the particle will be found in the eigenstate $\psi_{x'}(x)$  
is equal to $|c_{x'}|^2$. Recalling (\ref{appb4}), this again leads to
(\ref{appb1}).

Yet another way to understand (\ref{appb1}) and (\ref{appb2}) is through the theory of
{\em ideal} quantum measurements. The measured particle with the position $x$ entangles
with the measuring apparatus described by a pointer position $y$.
Thus, instead of (\ref{appb5}) we have the total wave function
\begin{equation}\label{appb6}
\Psi(x,y)=\sum_{x'} c_{x'} \psi_{x'}(x)  E_{x'}(y) ,
\end{equation}
where $E_{x'}(y)$ are normalized apparatus states that do not overlap
in the $y$-space. 
The main axiom (\ref{appb1}) now includes $y$ as well, so that
(\ref{appb1}) generalizes to $dP=|\Psi(x,y)|^2 d^4x \, dy$. 
Therefore, the probability that $y$ will have a value from the support of
$E_{x'}(y)$ is equal to $|c_{x'}|^2$, which again leads to (\ref{appb1}).
Eq.~(\ref{appb2}) emerges from (\ref{appb1})
as a conditional probability within the statistical ensemble of 
measurement outcomes. 

Of course, a real experiment is not ideal. If the departure from ideality is small, then
the distributions of measurement outcomes (\ref{appb1}) and  (\ref{appb2})
are good approximations. However, a real experiment may in fact be far from being ideal.
In particular, in a real experiment the functions $\psi_{x'}(x)$ (which are 
localized in both space and time) in (\ref{appb6}) may be replaced by functions
$\psi_{{\bf x}'}({\bf x},t)$ which are well localized in space, but very extended in time.
In such an experiment, (\ref{appb6}) is replaced by an entanglement of the form 
\begin{equation}\label{appb6'}
\Psi({\bf x},t,y)=\sum_{{\bf x}'} c_{{\bf x}'}(t) \psi_{{\bf x}'}({\bf x},t)  E_{{\bf x}'}(y) .
\end{equation}
For definiteness, $\psi_{{\bf x}'}({\bf x},t)$ may be taken to be time independent
and proportional to $\delta^3({\bf x}'-{\bf x})$ for 
$t\in [t_0-\Delta t/2, t_0+\Delta t/2]$, but vanishing for other values of $t$.
Now the measuring apparatus measures the space position ${\bf x}$ of the particle very well,
but time $t$ at which the particle attains this position remains uncertain with uncertainty
equal to $\Delta t$. 
Instead of (\ref{appb2}), the measured distribution of particle positions is now
\begin{equation}\label{appb2'}
 dP_{(3)}=\rho({\bf x}) \, d^3x ,
\end{equation}
where
\begin{equation}\label{appb2''}
 \rho({\bf x})=\frac{ \displaystyle\int_{t_0-\Delta t/2}^{t_0+\Delta t/2} 
dt \, |\psi({\bf x},t)|^2 }{N} ,
\end{equation}
and $N$ is the normalization factor chosen such that $\int d^3x \, \rho({\bf x}) =1$.
Namely, the probability that $y$ will have a value from the support of
$E_{{\bf x}}(y)$ is given by (\ref {appb2'}).

We also stress that the difference between (\ref{appb2}) and (\ref{appb2'})
is irrelevant in most practical situations. Namely, in actual experiments
one usually deals with wave functions that are well approximated by 
energy eigenstates with a trivial time dependence proportional to
$e^{-iEt}$. Consequently, $|\psi|^2$ is almost independent on $t$,
which makes (\ref{appb2}) and (\ref{appb2'}) practically indistinguishable. 

\subsection{Bohmian interpretation}

Now let us assume that particles have trajectories
\begin{equation}\label{appb7}
\frac{dX^{\mu}(s)}{ds}=v^{\mu}(X(s)),
\end{equation}
where $v^{\mu}(x)$ is such that
\begin{equation}\label{appb7'}
\partial_{\mu}(|\psi|^2 v^{\mu})=0 .
\end{equation}
Are such trajectories compatible with the probabilistic predictions studied
in Sec.~\ref{secAPPB1}? Clearly, they are compatible with (\ref{appb1})
because
\begin{equation}\label{appb8}
\frac{\partial |\psi|^2}{\partial s}+ \partial_{\mu}(|\psi|^2 v^{\mu})=0 .
\end{equation}
But are they compatible with (\ref{appb2})?
To answer this question, it is useful to eliminate $s$ from (\ref{appb7})
by introducing velocities
\begin{equation}\label{appb9}
\frac{dX^i}{dx^0}=\frac{v^i}{v^0}\equiv u^i , \;\; {\rm for} \;\; i=1,2,3 .
\end{equation}
In general, (\ref{appb7'}) implies that
\begin{equation}\label{appb10}
\frac{\partial |\psi|^2}{\partial t}+ \partial_i(|\psi|^2 u^i) \neq 0 .
\end{equation}
Inequality (\ref{appb10}) shows that, {\it a priori}, (\ref{appb7}) is not compatible with
(\ref{appb2}) (see also \cite{durr96}). 

Nevertheless, the compatibility restores when the theory of 
quantum measurements is also taken into account. Namely,  
we extend (\ref{appb7}) and (\ref{appb7'}) such that $Y(s)$ also
satisfies a Bohmian equation of motion compatible with  
$dP=|\Psi(x,y)|^2 d^4x \, dy$.
For ideal quantum measurements, (\ref{appb6}) implies that 
the probability that $Y$ will take a value from the support of
$E_{x'}(y)$ is equal to $|c_{x'}|^2$. 
This means that the statistics of measurement outcomes is given by
(\ref{appb1}). Thus, (\ref{appb2}) emerges from (\ref{appb1})
as a conditional probability within the statistical ensemble of
measurement outcomes. 
Similarly, in a more realistic measurement based on (\ref{appb6'}),
the probability that $Y$ will take a value from the support of
$E_{{\bf x}}(y)$ is given by (\ref{appb2'}).

Thus we see that the space probability density (\ref{appb2})
does not necessarily need to be correct. Instead, the space probability density depends
on how exactly it is measured.
However, the important points are (i) that the space probability density
can in principle be predicted from the fundamental axiom (\ref{appb1})
when the measuring procedure is well defined, and (ii) that
the probabilistic predictions of the Bohmian interpretation agree with 
those of the ``standard'' (pure probabilistic) interpretation for ideal measurements based on 
(\ref{appb6}), as well as for more realistic measurements based on (\ref{appb6'}).

\end{document}